\documentstyle[multicol,aps,epsfig]{revtex}
\begin{document}
\title{Comments on "Dimerization Induced by the RKKY Interaction" }
\maketitle
\begin{figure}[t]
\vspace{-2mm}
\centerline{
\psfig{file=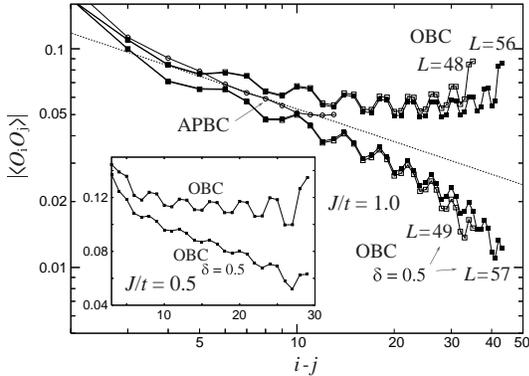,width=7cm}
}
\caption{ $|\langle O_i O_j \rangle |$ 
as a function of $|i-j|$ at $J\!=\!1$ under OBC($i\!=\!36,44$ for 
$L\!=\!48\!-\!49,56\!-\!57$, respectively) 
and APBC($L\!=\!32$). 
The dotted line proportional to $|i\!-\!j|^{-0.5}$ is a guide for the eye.
The inset shows the results at $J\!=\!0.5$. }
\label{fig1}
\vspace{-2mm}
\end{figure}
\begin{figure}[t]
\vspace{-2mm}
\centerline{
\psfig{file=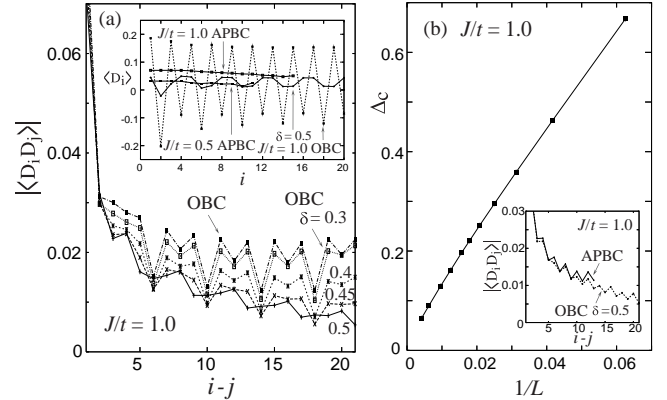,width=8.5cm}
}
\caption{(a) $|\langle D_i D_j \rangle |$ at $J\!=\!1$ 
under several OBC's. 
The inset shows $\langle D_j\rangle$ under OBC and APBC. 
(b) Charge gap, $\Delta_c$, as a function of $L^{-1}$. 
(The inset shows $|\langle D_j D_{j+l} \rangle |$ of APBC and 
the system-average of $|\langle D_j D_{j+l} \rangle |$ under 
OBC with $\delta$=0.5.)
}
\label{fig2}
\vspace{-2mm}
\end{figure}

\begin{multicols}{2}
\narrowtext
Recently, Xavier {\it et al.} claimed the existence of an insulating 
spin dimer state in the one-dimensional Kondo lattice model 
at quarter-filling amidst the paramagnetic metallic 
phase\cite{xavier}. 
Such dimer state is characterized by $q\!=\!\pi$ oscillations 
of the nearest-neighbor localized spin correlation, 
$\langle D_i \rangle \!\!=\!\!\langle S_i \!\cdot \!S_{i+1}\rangle$, 
under open boundary conditions(OBC) 
in their density matrix renormalization group(DMRG) analysis\cite{white}, 
where they increased the system size up to $L\!=\!120$ keeping 
$m\!=\!800$ states per block. 
In this comment we show that such oscillations of $\langle D_i\rangle$
do not persist in the bulk limit and that the correlation function, 
$\langle D_i D_j \rangle$, decays algebraically at long distances.
\\
We first present in Fig.~\ref{fig1} the value of dimer-dimer correlation function, 
$|\langle O_i O_j \rangle|$, $O_i\!\equiv\!D_i\!-\!D_{i+1}$ obtained by
the DMRG under several boundary conditions(BC). 
As for the OBC it decreases first, and then increases near the 
boundary of the system (i.e. $|i\!-\!j|\!\sim \!L$) as expected 
from the report of Xavier {\it et al.}. 
However, when we add an extra site to the system 
and put the potentials, $\mu\delta$ and $\mu(1-\delta)$,
at the end sites, it decays significantly fast towards zero. 
Here, we control the potential $\mu$ 
to keep the density of the electrons at quarter-filling.
We find that with increasing system size, these two results 
become close together 
and seem to asymptotically approach the line, 
$|\langle O_i O_j\rangle| \propto |i\!-\!j|^{-0.5}$. 
This power law decay 
%%towards $\lim_{|i-j|\rightarrow \infty} |\langle O_i O_j\rangle|\sim 0$ 
is confirmed by the results obtained under the anti-periodic 
boundary conditions(APBC) as shown in the figure. 
%%%%%%%%%% Fig. 1 %%%%%%%%
%%%%%%%%%%%%%%%%%%%%%%%%
We next discuss the effect of BC in detail. 
Figure~\ref{fig2}(a) shows $|\langle D_i D_j \rangle|$ when the BC 
is modified from $\delta\!=\!0$ to 0.5. 
The correlation function varies smoothly from the one at 
$\delta$=0, where $\delta$=0 reproduces the result of the usual OBC. 
The corresponding $\langle D_j \rangle$ given in the inset of 
Fig.~\ref{fig2}(a) shows that the large oscillation of $\langle D_j \rangle$
characteristic of the dimer state disappears at $\delta$=0.5. 
It should be noted that the energy density away from the boundaries
is the same between $\delta$=0 and 0.5 within the truncation error.
These results indicate that the spin structure is sensitive to the BC 
when the translational invariance(TI) is violated. 
Actually, $\langle D_i\rangle$ under APBC and OBC at $\delta$=0.5 
are almost $i$-independent, i.e. the system has TI. Therefore
$|\langle D_i D_j \rangle|$ at $\delta\!=\!0.5$ 
is almost the same as that obtained under
the APBC as shown in the inset of Fig.~\ref{fig2}(b), 
where we have removed small $i$-dependence in $\langle D_i\rangle$ 
by taking an average over the system. 
These results indicate that the dimer structure is 
induced by the BC which breaks the TI,  
and it is not an intrinsic bulk property. 
%%%%%%%%%% Fig. 2 %%%%%%%%
%%%%%%%%%%%%%%%%%%%%%%%%
The proper examination of the present issue requires severe numerical 
accuracy. We calculated the system size dependence up to $L\!=\!240$ 
taking a maximum of $m\!=\!1100$ 
to confirm the absence of a dimer long range order. 
In fact, more than $m\!=\!600$ states are necessity even at $L$=32 for $J$=1, 
and we found that one can no longer get enough reliable results for 
$L\!>\!80$ at $J\!=\!0.5$ even with $m\!=\!1100$.
We also calculate the charge gap, $\Delta_c(L)$, 
at $J\!=\!1$ where the numerical accuracy is well controlled by $m$. 
As shown in Fig.~\ref{fig2}(b), 
$\Delta_c(L)$ behaves slightly convex upwards as a function of $L^{-1}$ 
towards the bulk limit, 
$\Delta_c(\infty) \leq 10^{-2}$, and the system is expected
to be metallic. 
\par
In conclusion, power law decay of the dimer-dimer correlation 
and the possible absence of a charge gap are confirmed. 
At present, there is no evidence which denies the metallic property
of the ground state at quarter-filling characterized by a Tomonaga-Luttinger liquid. 
\vspace{2mm}
\\
{\small 
{\bf Naokazu Shibata}\\
{\it Dept. Basic Science, University of Tokyo, Tokyo 153-8902, Japan.}\\
{\bf Chisa Hotta}\\
{\it Dept. Physics and Mathematics, Aoyama-Gakuin University, Kanagawa 229-8558, Japan.}\\
}
%%%%%%%%%  REFER  %%%%%%%%%%%%%%%%%%%%%%%%%%%%%%%%%%%%%%%

%%%%%%%%%%%%%%%%%%%%%%%%%%%%%%%%%%%%%%%%%%%%%%%%%%%%%%%%%%
\end{multicols}
\end{document}